\documentclass [onecolumn,amsmath,amssymb,aps]{revtex4}
\usepackage {graphicx}% Include figure files
\usepackage {dcolumn}% Align table columns on decimal point
\usepackage {bm}% bold math

\begin {document}

\section {Introduction}

The magnetic and phonon pairing mechanisms dominate in the origin of high-temperature superconductivity (HTSC) \cite {Maksimov}. As a rule, the $t - t' - t'' - J$ model is used for describing the magnetic mechanism. Analysis of the electron structure of HTSC cuprates based on the multiband $pd$-model in the strong electron correlation (SEC)  limit proves that the effective low-energy model for cuprates is defined by the $t - t' - t'' - J^*$ model differing from the $t - t' - t'' - J$  model in the addition of three-center correlated hoppings (the effect of these hoppings on the  $T_c $ value was noted in \cite {Valkov}). The parameters of the $t - t' - t'' - J^*$ Hamiltonian have recently been calculated {\it ab initio} for $La_{2 - x} Sr_x CuO_4 $ taking into account SECs in the LDA + GTB scheme (local density approximation + generalized tight binding technique) \cite {Korshunov}. The Hamiltonian for the electron-phonon interaction (EPI) was also derived taking into account SECs. Explicit dependences of the EPI matrix elements on the transferred and input wave vectors were derived for three modes most strongly interacting with electrons \cite {Ovchinnikov}. Here, in the mean field approximation, we calculate the superconducting transition temperature as a function of the doping level for a superconductor with the $d_{x^2  - y^2 } $ symmetry. We show that the magnetic mechanism provides too high values for $La_{2 - x} Sr_x CuO_4 $, while the phonon contribution lowers the superconducting transition temperature. The phonon-induced suppression of $T_c $ is associated with the predominant contribution from the breathing mode, for which the EPI is the strongest for large momentum transfers and changes the sign of the $d_{x^2  - y^2 }$ symmetry order parameter.

\section {The effect of the electron-phonon interaction on the superconducting order parameter} 

In the $X$-operator representation, the electron-phonon interaction in the strong correlation mode can
be described by the effective low-energy Hamiltonian \cite {Ovchinnikov} $
H_{tot}  = H_{el}  + H_{el - ph - el}$, where $H_{el}$ is the Hamiltonian of the $t - t' - t'' - J^*$ model and $H_{el-ph-el}$ takes into account the
interaction of electrons via the emission and absorption of phonons:
\begin {equation}
H_{el - ph - el}  = \sum\limits_{{\bf{kk'q}}} {\sum\limits_{\sigma \sigma '} {V_{{\bf{kk'q}}} X_{{\bf{k + q}}}^{\sigma 0} X_{{\bf{k}}' - {\bf{q}}}^{\sigma '0} X_{{\bf{k'}}}^{\;0\sigma '} X_{\bf{k}}^{\;0\sigma } } }, 
\label {eq1}
\end {equation}
where $X_{{\bf{k}} + {\bf{q}}}^{0\sigma } $ is the creation operator for a hole with spin $\sigma $ and momentum ${\bf{k}} + {\bf{q}}$ and $V_{{\bf{kk}}'{\bf{q}}} $ is the matrix element of the effective interaction, which has the same structure as in the Frolich theory \cite {Frohlich}. In contrast to the theory of weakly correlated electrons, $V_{{\bf{kk}}'{\bf{q}}} $ depends on the band filling factor $F_{(0\sigma )}$ and, hence, on the doping level, temperature, and magnetic field:
\begin {equation}
V_{{\bf{kk'q}}}  = \sum\limits_\nu  {\frac{{g_\nu  \left( {{\bf{k}},{\bf{q}}} \right)g_\nu  \left( {{\bf{k}}',{\bf{ - q}}} \right)\omega _{{\bf{q}},\nu } }}{{\left( {\varepsilon _{\bf{k}}  - \varepsilon _{{\bf{k}} + {\bf{q}}} } \right)^2 F_{(0\sigma )} ^2  - \omega _{{\bf{q}},\nu }^2 }}} .
\label {eq2}
\end {equation}
Here, $g_\nu  \left( {{\bf{k}},{\bf{q}}} \right)$ is the matrix element of the interaction between an electron with initial momentum ${\bf{k}}$ and a phonon with momentum ${\bf{q}}$, $\omega _{{\bf{q}},\nu }$ is the vibrational frequency of mode $\nu$, and ${\varepsilon _{\bf{k}} }$ is the Fourier transform of the hopping integral. It should be noted that, in deriving the
effective EPI in the low-energy $tJ^*$ model, we disregarded interband excitations via the gap $E_{ct}$ with charge transfer, which are associated with lattice vibrations. The resulting Hamiltonian corresponds to the lower Hubbard subband of holes for electron-doped systems. The Hamiltonian for hole-doped cuprates with carriers in the upper Hubbard subband has an analogous structure \cite {Ovchinnikov}.

The effect of the EPI on the superconducting order parameter is considered in the generalized Hartree–
Fock approximation using the method of irreducible Green's functions for Hubbard operators \cite {Tyablikov, Plakida}. The
mean values $\left\langle {X_f^{\sigma \sigma } X_g^{\sigma '\sigma '} } \right\rangle$  appearing in this case can be
written taking into account static spin correlation functions $c_{\bf{q}}  = \left\langle {X_{\bf{q}}^{\sigma \bar \sigma } X_{\bf{q}}^{\bar \sigma \sigma } } \right\rangle  = \left\langle {S_{\bf{q}}^ +  S_{\bf{q}}^ -  } \right\rangle $ \cite {Oudovenkko}, which enables us to go beyond the scope of the generalized Hartree–Fock approximation. Kinematic correlation functions $\left\langle {X_{\bf{q}}^{\sigma 0} X_{\bf{q}}^{0\sigma } } \right\rangle $ were omitted, which did not change the pattern qualitatively: kinematic correlation functions are an order of magnitude smaller than spin correlation functions in the region, where electron correlations are significant (from weakly doped to optimally doped
compositions)  \cite {Dzebisashvili}. Finally, taking into account the fact that anomalous means $ B_{\bf{q}}  = \left\langle {X_{ - {\bf{q}}}^{0, - \sigma } X_{\bf{q}}^{0,\sigma } } \right\rangle $  for singlet pairings of the (s, d type) have the symmetry $B_{\bf{q}}  = B_{ - {\bf{q}}}$, we can write the order parameter in the form $\Delta _{\bf{k}}^{tot}  = \Delta _{\bf{k}}^{tJ*}  + \Delta _{\bf{k}}^{el - ph} $. The parameter $\Delta _{\bf{k}}^{tJ*} $ has the form, which is standard for the gap in the $tJ^*$ model \cite {Zaoetsev}:
\begin {equation}
\Delta _{\bf{k}}^{tJ*}  = \frac{1}{N}\sum\limits_{\bf{q}} {\left( { - \frac{4}{{1 + x}}t_{\bf{q}}  - \frac{{1 - x}}{{1 + x}}\left( {J_{{\bf{k + q}}}  + J_{{\bf{k}} - {\bf{q}}} } \right) - \frac{{4t_{\bf{k}} t_{\bf{q}} }}{{E_{ct} }} + \frac{{1 - x}}{{1 + x}} \cdot \frac{{t_{\bf{q}} ^2 }}{{E_{ct} }}} \right)} B_{\bf{q}} 
\label {eq3}
\end {equation}
where the first term is known to be determined by the kinematic mechanism \cite {Zaoetsev}, the second term corresponds
to exchange pairing renormalized by three-center interactions \cite {Valkov, Yushankhay}, and the third and fourth terms appear due also to three-center interactions. Further, we analyze the structure of $\Delta _{\bf{k}}^{el - ph} $:
\begin {equation}
\begin{array}{l}
 \Delta _{\bf{k}}^{el - ph}  = \frac{1}{N}\sum\limits_{\bf{q}} {\frac{{1 + x}}{4}\left( {V_{ - {\bf{q}}{\bf{,q}}{\bf{,q + k}}}  + V_{ - {\bf{q}}{\bf{,q}}{\bf{,q}} - {\bf{k}}} } \right)} B_{\bf{q}}  -  \\ 
  - \frac{1}{{N^2 }}\sum\limits_{{\bf{q}},{\bf{p}}} {\frac{3}{{2\left( {1 + x} \right)}}} \left( {V_{ - {\bf{q}}{\bf{,q}}{\bf{,p + k}}}  + V_{ - {\bf{q}}{\bf{,q}}{\bf{,p}} - {\bf{k}}} } \right)B_{\bf{q}} c_{{\bf{q - p}}} . \\ 
 \end{array}
\label {eq4}
\end {equation}
The first term describes the phonon pairing mechanism in the mean field theory and the second term is associated
with interference of the magnetic and phonon pairing mechanisms. It is important that the appearance of the contribution proportional to the product of the EPI constant ${V_{ - {\bf{q}}{\bf{,q}}{\bf{,p + k}}} }$ and the spin correlation function $c_{{\bf{q - p}}}$ is a manifestation of the spin liquid effects. In the region of strong doping, where spin correlations can be ignored, the spin-liquid contribution vanishes, and the spin-liquid effect enhances the EPI in the region of
weak doping.

\section {The effect of the electron-phonon interaction on the superconducting transition temperature in $La_{2-x}Sr_xCuO_4$. Conclusion.}

Let us estimate the effect of the EPI on the superconducting transition temperature in $La_{2 - x} Sr_x CuO_4 $. To analyze the EPI constant, let us consider the optical modes most strongly interacting with electrons in the $CuO_{2}$ plane \cite {Pintschovius, Pint, Nunner, Song, Kulic, Bulut}, namely, the longitudinal breathing mode (vibrations of oxygen ions in the $CuO_{2}$ plane, which deform the $Cu–O$ bond), the apical breathing mode (vibrations of apical oxygen ions deforming the
$Cu–O$ bond along the $c$ axis), and the buckling mode (vibrations of oxygen ions in the $CuO_{2}$ layer across the $Cu–O$ bond). In the general case, the matrix element $g_\nu  \left( {{\bf{k}},{\bf{q}}} \right)$ of the interaction can be represented as the sum of the diagonal and off-diagonal (with respect to lattice sites) contributions; in this sum, only the latter contribution depends on the initial electron momentum ${\bf{k}}$. For all above-mentioned modes, the dependence of matrix
elements on wave vectors was determined explicitly. For example, for the breathing mode $\nu  = 1$, we have
\begin {equation}
\begin{array}{l}
\ g_{dia}^{\left( 1 \right)} \left( q \right) = \frac{{2i\upsilon _{dia}^{\left( 1 \right)} }}{{\sqrt {2M_O \omega _{q,1} } }}\left( {e_x^{\left( {O_x } \right)} \sin \frac{{q_x a}}{2} + e_x^{\left( {O_y } \right)} \sin \frac{{q_y a}}{2}} \right), \\ 
 \ g_{off}^{\left( 1 \right)} \left( {{\bf{k}},q} \right) = \frac{{8i\upsilon _{off}^{\left( 1 \right)} }}{{\sqrt {2M_O \omega _{{\bf{q}},1} } }}\left[ {e_x^{\left( {O_x } \right)} \sin \frac{{q_x a}}{2} + e_x^{\left( {O_y } \right)} \sin \frac{{q_y a}}{2}} \right]\left[ {\gamma \left( {\bf{k}} \right) + \gamma \left( {{\bf{k + q}}} \right)} \right] \\ 
 \end{array}
\label {eq5}
\end {equation}
where ${\upsilon _{dia}^{\left( 1 \right)} }$ and ${\upsilon _{off}^{\left( 1 \right)} }$ are the parameters of the diagonal
and off-diagonal EPIs, respectively; $M_{O}$ is the mass of oxygen atoms; $e_{\alpha ,\nu } $ is the polarization vector; and $\gamma \left( {\bf{q}} \right) = \left( {\cos q_x a + \cos q_y a} \right)/2$. For the buckling mode $\nu  = 2$, we have
\begin {equation}
\begin{array}{l}
\ g_{dia}^{\left( 2 \right)} \left( q \right) = \frac{{2\upsilon _{dia}^{\left( 2 \right)} }}{{\sqrt {2M_O \omega _{{\bf{q}},2} } }}\left[ {e_z^{\left( {O_x } \right)} \cos \frac{{q_x a}}{2} + e_z^{\left( {O_y } \right)} \cos \frac{{q_y a}}{2}} \right], \\ 
 g_{off}^{\left( 2 \right)} \left( {{\bf{k}},q} \right) = \frac{{2\upsilon _{off}^{\left( 2 \right)} }}{{\sqrt {2M_O \omega _{{\bf{q}},2} } }}\left[ {e_z^{\left( {O_x } \right)} \cos \left( {k_x  + \frac{{q_x }}{2}} \right)a + e_z^{\left( {O_y } \right)} \cos \left( {k_y  + \frac{{q_x }}{2}} \right)a} \right]. \\
\end{array}
\label {eq6}
\end {equation}
Finally, for the apical breathing mode $\nu  = 3$, we have
\begin {equation}
g_{dia}^{\left( 3 \right)} \left( {\bf{q}} \right) = \frac{{g_{dia,m}^{\left( 3 \right)} }}{{\sqrt {2M_O \omega _{{\bf{q}},3} } }}{\rm{e}}_z^{\left( {O_{ap} } \right)} ,{\rm{  }}g_{off}^{\left( 3 \right)} \left( {{\bf{k}},{\bf{q}}} \right) = 0.
\label {eq7}
\end {equation}
An analogous wave-vector dependence of the EPI was obtained for $g_{dia}^{\left( 1 \right)} \left( q \right)$
, $g_{off}^{\left( 1 \right)} \left( {{\bf{k}},q} \right)$, and $g_{dia}^{\left( 2 \right)} \left( q \right)$ (see, e.g., \cite {Ishihara, Sherman}). Using Eqs. (5-7), we analyze the superconducting state with the symmetry $d_{x^2  - y^2 } $. In this case, the expression for the gap has the form:
\begin {equation}
\Delta _{\bf{k}}^{\left( {d_{x^2  - y^2 } } \right)}  =  - \frac{1}{N}\varphi _{\bf{k}} \sum\limits_{\bf{q}} {\left( {J\frac{{1 - x}}{{1 + x}}} \right.} \left. { + G\frac{{3\left( { - c_{01} } \right) + 0.5\left( {1 + x} \right)^2 }}{{2\left( {1 + x} \right)}}\theta \left( {\left| {\xi _{\bf{q}}  - \mu } \right| - \omega _D } \right)} \right)B_{\bf{q}} \varphi _{\bf{q}} ,
\label {eq8}
\end {equation}
where $
G{\rm{ }} = {\rm{ }}\frac{{\upsilon _{dia,\nu  = 2}^2 }}{{\omega _{\nu  = 2} }} - \frac{{\upsilon _{dia,\nu  = 1}^2 }}{{\omega _{\nu  = 1} }}
$, ${\rm{ }}\varphi _{\bf{k}}  = \left( {\cos k_x  - \cos k_y } \right)$, and $\theta \left( x \right) = 0 $ for $x > 0$, $\theta \left( x \right) = 1 $ for $x < 0$. The $\theta$-function appears because the phonon contribution is significant only in a narrow layer $ \sim \omega _D $ near the Fermi surface. The coupling constant $\alpha _{tot}  = \alpha _{tJ*}  + \alpha _{el - ph} $ appearing in the braces in Eq. (8) is described by the magnetic pairing mechanism in the $tJ^*$ model, which is renormalized by the EPI (Fig. 1). With increasing $x$ (number of carriers), the magnetic and spin-liquid contributions decrease, while the phonon contribution increases. The competition of the spin-liquid and phonon contributions leads to an
increase in the EPI in the weak-doping region. It should be noted that the EPI matrix elements appear in Eq. (8) in the form of the combined parameter $G$ whose sign determines whether the EPI increases or reduces the total coupling constant (see Fig. 1a, 1b). According to Eq. (8), the contribution of the apical breathing mode in the channel vanishes in this case, the buckling mode facilitates electron pairing, and the breathing mode reduces the pairing potential. Equation (8) was derived taking into account the explicit dependence of matrix elements on wave vectors ${\bf{k}}$ and ${\bf{q}}$. The origin of contributions from various vibrations to the EPI can be explained as follows. Obviously, for the $s$ type order parameter, all vibrational modes facilitate electron pairing: the EPI is strongest when an electron near the Fermi surface returns to this surface after interaction with a phonon transferring momentum ${\bf{q}}$. In the case of the $s$-type gap, the interaction of electrons with
phonons with any momentum transfer does not change the sign of the order parameter (Fig. 2a). For a superconducting state with the $d_{x^2  - y^2 } $ symmetry, the interaction between electrons and phonons for large values of ${\bf{q}}$ changes the sign of the order parameter and, hence, lowers the pairing potential (Fig. 2b). Thus, the breathing mode having the interaction peak for large momentum transfers ${\bf{q}}$ makes a negative contribution to the coupling constant of the $d_{x^2  - y^2 } $ type. The interaction with the buckling mode, which is strongest for small ${\bf{q}}$ values, increases $\alpha _{tot} $. (In the case of $s$-type pairing, the coupling constant is proportional to the sum of the matrix elements of all modes, and the EPI increases $T_c$.)

\begin {figure}
\includegraphics[width=0.25\linewidth]{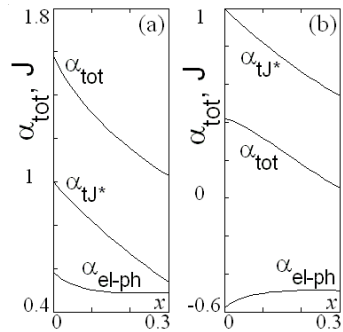}
\caption {Effective coupling constant $\alpha _{tot} $ of superconducting pairing determined by the sum of the magnetic $\alpha _{tJ*}$ and phonon $\alpha _{el - ph}$ mechanisms for the EPI parameters $G/J=1$ (a) and $G/J=-1$ (b).}
\end {figure}

Let us now consider the self-consistent equation determining the superconducting transition temperature,
\begin {equation}
1 = \frac{1}{N}\sum\limits_{\bf{q}} {\left\{ {\frac{{1 - x}}{2}J + \left( {\frac{{3\left( { - c_{01} } \right)}}{{4\left( {1 + x} \right)}} + \frac{{\left( {1 + x} \right)}}{8}} \right)\theta \left( {\left| {\xi _{\bf{q}}  - \mu } \right| - \omega _D } \right)G} \right\}}  \times \frac{{\left( {\cos q_x  - \cos q_y } \right)^2 }}{{\xi _{\bf{q}}  - \mu }}\tanh \left( {\frac{{\xi _{\bf{q}}  - \mu }}{{2T_c }}} \right)
\label {eq9}
\end {equation}
where ${\xi _{\bf{q}} }$ is the normal phase dispersion taking into account spin correlation functions and three-center
interactions \cite {Dzebisashvili} and $\mu $ is the chemical potential. Equation (9) was solved together with the equation for the chemical potential for hole concentration $1 + x$, which corresponds to $La_{2-x}Sr_xCuO_4$. Figure 3 shows the results of numerical solution. The quasiparticle spectrum in the normal phase is described without using any fitting parameters, because all parameters of the $t-t'-t''-J^*$ model were obtained for $La_{2-x}Sr_xCuO_4$ in the LDA + GTB scheme, which combines the {\it ab initio} and model approaches \cite {Korshunov}. The spin correlation functions were calculated self-consistently in \cite {Sherman}. The only free parameter in this approach is the effective EPI constant $G$. In this case, the position of $x_{opt}$ corresponding to the given parameters is virtually independent of $G$ and is in good agreement with the experiment.

\begin {figure}
\includegraphics[width=0.3\linewidth]{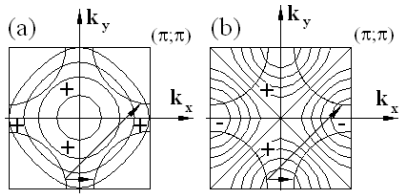}
\caption {Interaction of electrons with phonons does not change the sign of the order parameter for $s$ pairing (a) and changes the sign for the $d_{x^2  - y^2 } $ pairing (b).}
\end {figure}

As is seen in Fig. 3, the magnetic pairing mechanism gives (middle line) too high $T_c$ values as compared to experiment. The effect of the EPI is determined by the combined parameter $G$. For $G > 0$ and $G < 0$, the EPI respectively (upper line) increases and (lower line) decreases the superconducting transition temperature associated with the magnetic mechanism. It should also be noted that the inclusion of hoppings to the second and third coordination spheres in deriving Eqs. (8) and (9),
as well as the off-diagonal part of the EPI, leads to the appearance of higher harmonics $\varphi _m \left( {\bf{k}} \right) = \cos \left( {mk_x } \right) - \cos \left( {mk_y } \right)$ in the order parameter. For fixed ratios $t/t'$ and $t/t''$, higher harmonics do not change the position of the peak on the concentration dependence $T_c(x)$ \cite {Val}.

Let us analyze the sign of the effective constant $G$ in $La_{2-x}Sr_xCuO_4$. In most models for studying the EPI in $p$-type cuprates, only $CuO_2$ planes are considered. In such an approach, in view of the symmetry of vibrations, the interaction of an electron with the buckling mode appears only due to anharmonism \cite {Song}. A small contribution linear in oxygen displacements appears in the corrugated $CuO_2$ layer due to orthorhombic distortions. The inclusion of the apical oxygen atom in more realistic models leads to a linear contribution to the EPI, which is small in the hybridization parameter \cite {Ovchinnikov}. Calculation of the EPI matrix elements (e.g., in \cite {Song}) also shows that the interaction of an electron with the breathing mode is stronger than with the buckling mode. The above consideration indicates that the effective constant $G$ is negative in $p$-type cuprates. The negative value of constant $G$ is also confirmed by analysis of kinks. According to the ARPES (angle-resolved photoemission) data, the manifestation of a kink near the nodal point $\left( {\pi /2;\pi /2} \right)$
is much stronger that at the antinodal point $\left( {\pi;0} \right)$ \cite {Lanzara, Cuk}. It was concluded \cite {Cuk, Ovchinnikov} that kinks at the nodal and antinodal points appear due to the interactions of electrons with the breathing and
buckling modes, respectively. Thus, depending on the order parameter symmetry, the EPI leads to the following results: the EPI facilitates pairing in $s$-type superconductors, while the $T_c$ value for the $d_{x^2  - y^2 } $ type may both increase and decrease depending on the relations between the EPI matrix elements and various modes. In particular, for $La_{2-x}Sr_xCuO_4$, the EPI lowers the superconducting transition temperature associated with the magnetic pairing mechanism.
\begin {figure}
\includegraphics[width=0.35\linewidth]{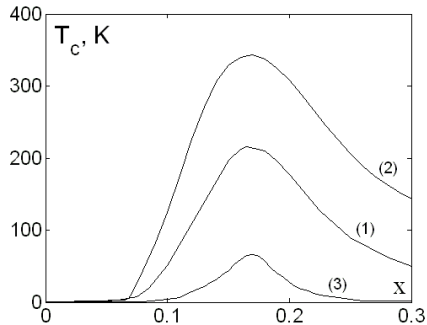}
\caption {Concentration dependence of the superconducting transition temperature for the effective EPI parameters $G/J$
indicated near the curves.}
\end {figure}

\begin {acknowledgments} We are grateful to V.I. Zinenko, V.V. Val'kov,
D.M. Dzebisashvili, M.M. Korshunov, Yu.A. Izyumov,
and N.M. Plakida for fruitful discussions. This study
was supported by the Presidium of the Russian Academy
of Sciences (program "Quantum Microphysics"),
the Siberian Division of the Russian Academy of Sciences
(complex integration project no. 3.4), and the
Russian Foundation for Basic research (project no. 06-
02-16100). E.I.Sh. acknowledges the support of the
Dynasty Foundation, the International Center of Fundamental
Physics in Moscow, and the Siberian Division
of the Russian Academy of Sciences (Lavrentiev Competition
of Young Scientists' Projects). 
\end {acknowledgments}


\begin {thebibliography}{23}
\bibitem {Maksimov} E. G. Maksimov, Usp. Fiz. Nauk 170, 1033 (2000) [Phys. Usp. 43, 965 (2000)].
\bibitem {Valkov} V. V. Val'kov, T. A. Val'kova, D. M. Dzebisashvili, and S. G. Ovchinnikov, Pis'ma Zh. Eksp. Teor. Fiz. 75, 450 (2002) [JETP Lett. 75, 378 (2002)].
\bibitem {Korshunov} M. M. Korshunov, V. A. Gavrichkov, S. G. Ovchinnikov, et al., Phys. Rev. B 72, 165104 (2005).
\bibitem {Ovchinnikov}  S. G. Ovchinnikov and E. I. Shneyder, Zh. Eksp. Teor. Fiz. 128, 974 (2005) [JETP 101, 844 (2005)].
%
\bibitem {Frohlich} H. Frohlich, Phys. Rev. 79, 845 (1950). 
\bibitem {Tyablikov} S. V. Tyablikov, Methods in the Quantum Theory of Magnetism, 2nd ed. (Nauka, Moscow, 1975; Plenum, New York, 1967). 
\bibitem {Plakida} N. M. Plakida, V. Yu. Yushankhay, and I. V. Stasyuk, Physica C (Amsterdam) 162, 787 (1989).
\bibitem {Oudovenkko} N. M. Plakida and V. S. Oudovenkko, Phys. Rev. B 59, 11949 (1999).
\bibitem {Dzebisashvili} V. V. Val'kov and D. M. Dzebisashvili, Zh. Eksp. Teor. Fiz. 127, 686 (2005) [JETP 100, 608 (2005)]. 
\bibitem {Zaoetsev} R. O. Zaoetsev and V. A. Ivanov, Pis'ma Zh. ?ksp. Teor. Fiz. 46, 140 (1987) [JETP Lett. 46, 116 (1987)]. 
%
\bibitem {Yushankhay} V. Yu. Yushankhay, G. M. Vujicic, and R. B. Zakula, Phys. Lett. A 151, 254 (1990).
\bibitem {Pintschovius} L. Pintschovius and M. Braden, Phys. Rev. B 60, 15039 (1999).
\bibitem {Pint} L. Pintschovius, Phys. Status Solidi B 242, 30 (2005).
\bibitem {Nunner} T. S. Nunner, J. Schmailian, and K. N. Bennemann, Phys. Rev. B 59, 8859 (1999).
%
\bibitem {Song} J. Song and J. F. Annett, Phys. Rev. B 51, 3840 (1995).
\bibitem {Kulic} M. L. Kulic and O. V. Dolgov, Phys. Status Solidi B 242, 151 (2005).
\bibitem {Bulut} N. Bulut and D. J. Scalapino, Phys. Rev. B 54, 14971 (1996).
%
\bibitem {Ishihara} S. Ishihara and N. Nagaosa, Phys. Rev. B 69, 144520 (2004).
\bibitem {Sherman} A. Sherman and M. Schreiber, Eur. Phys. J. B 32, 11 (2003).
\bibitem {Val} V. V. Val'kov and D. M. Dzebisashvili, Pis'ma Zh. Eksp. Teor. Fiz. 77, 450 (2003) [JETP Lett. 77, 381 (2003)].
%
\bibitem {Lanzara}  A. Lanzara, P. V. Bogdanov, X. J. Zhou, et al., Nature 412, 510 (2001).
\bibitem {Cuk} T. Cuk, D. H. Lu, X. J. Zhou, et al., Phys. Status Solidi B 242, 11 (2005).
\end {thebibliography}
\end {document}